\documentclass[prd,onecolumn,preprintnumbers,nofootinbib]{revtex4}
\usepackage{graphicx}

\usepackage[plainpages=false, colorlinks=true, anchorcolor=blue, linkcolor=blue, citecolor=blue, bookmarks=false]{hyperref}
\usepackage{color}
\usepackage{float}
\usepackage{amsfonts,amsmath,amssymb}
\usepackage{natbib}
\usepackage{array}
\begin{document}
\newcolumntype{P}[1]{>{\centering\arraybackslash}p{#1}}
\pdfoutput=1
\newcommand{\jcap}{JCAP}
\newcommand{\araa}{Annual Review of Astron. and Astrophys.}

\newcommand{\aj}{Astron. J. }
\newcommand{\mnras}{MNRAS}
\newcommand{\apjl}{Astrophys. J. Lett.}
\newcommand{\apjs}{Astrophys. J. Suppl. Ser.}
\newcommand{\aap}{Astron. \& Astrophys.}
\renewcommand{\arraystretch}{2.5}
\title{Search for spatial coincidence between IceCube neutrinos and  radio pulsars}
\author{Vibhavasu \surname{Pasumarti}}
\altaffiliation{E-mail:ep20btech11015@iith.ac.in}

\author{Shantanu \surname{Desai}}
\altaffiliation{E-mail: shntn05@gmail.com}

\begin{abstract}
We search for a spatial association between radio pulsars and   ultra-high energy neutrinos  using the publicly available IceCube point source neutrino events catalog. For this purpose we  use the unbinned maximum likelihood method to search for a statistically significant excess from each of the pulsars in the ATNF catalog. We do not find any pulsars with detection significance much higher than that expected   from a Gaussian distribution,  Therefore, we conclude that none of the currently known pulsars  contribute to the diffuse neutrino flux detected by IceCube.
\end{abstract}

\affiliation{Dept  of Physics, IIT Hyderabad,  Kandi, Telangana-502284, India}

\maketitle
\section{Introduction}
\label{sec:intro}

The IceCube collaboration has detected a diffuse flux of astrophysical neutrinos in the TeV-PeV energy range, whose origin is still a mystery ~\cite{IceCube13}. This diffuse flux of neutrinos may also hold the key to unveil the source of ultra high energy cosmic rays~\cite{Olinto}. With the exception of some point sources
such as NGC 1068, TXS 0506+056, PKS 1424+240, the majority of IceCube events cannot be attributed to any astrophysical source~\cite{IceCubedata}.
A large number  of extragalactic astrophysical sources have been considered for their origin including blazars and other types of AGNs, star-forming galaxies, FRBs, GRBs, galaxy clusters, other ancillary extragalactic sources in Fermi-LAT catalog, etc.   
However, the expected contribution of blazars from the 2nd Fermi-LAT AGN catalog to the diffuse  neutrino flux cannot be more than 27\%~\cite{IceCubeAGN}.
Furthermore, no statistically significant spatial  with ultrahigh energy cosmic rays, believed to be extragalactic
have been seen~\cite{IceCubeuhecr}.
Although, isotropy studies of neutrino arrival directions indicate that any galactic component must be sub-dominant~\cite{IceCubeptsource}, it is still plausible that the galactic contribution could be up-to 10-20\%~\cite{Palladino}. If one posits  a 100\% extragalactic origin  for the neutrino flux there is an underlying tension when trying to fit the Fermi-LAT data~\cite{Abbasi2021}.
A recent review concludes that the neutrino flux component between 10-100 TeV could be of galactic origin~\cite{Troitsky}. 

Therefore, for the above reasons, a large number of works have explored possible contributions from galactic sources to the IceCube diffuse neutrino flux. This includes individual sources such as supernova remnants, X-ray binaries, pulsar wind nebulae, open clusters, LHAASO sources, as well as a diffuse galactic component due to cosmic ray interactions with the interstellar medium ~\cite{Lunardini,Marfatia,Icecubegalactic,IceCubeXRB,IceCubePWN,Kovalev22,LHAASO,Li22}. It is  also important to look for  coincidences sources which have hitherto not been considered.

In this work, we search for a statistically significant angular correlation between IceCube neutrinos and radio pulsars. Pulsars are rotating neutron stars, which emit pulsed radio emissions with periods ranging from  milliseconds to a few seconds  with  magnetic fields ranging from $10^8$ to $10^{14}$ G~\cite{handbook,Reddy}. Pulsars are known to be wonderful  laboratories for a whole slew of topics in Physics and Astronomy~\cite{Blandford92} from stellar evolution~\cite{Lorimer08} to dark matter~\cite{Kahya}. We note that pulsars and magnetars   have been known to be promising sources of high energy neutrinos for  a long time~\citep{Helfand,Fang15,Fang16}. Some  theoretical models have predicted steady state emissions from some  pulsars, with  an  event rate of about $\sim$ 5 events/km$^2$/year~\cite{Burgio,Bhadra}. Furthermore, a $2.6\sigma$ excess of GeV neutrinos from underground detectors has also been seen from one pulsar, viz PSR B1509-58~\cite{Desai22}. Neutrinos with GeV energies  may have been observed from the protoneutron star within one year of  SN 1987A explosion~\cite{Oyama}.
Furthermore, a large number of  mainly millisecond pulsars have been  detected by the Fermi-LAT telescope of which some are yet to be detected in radio~\cite{Acero}.  Pulsar   have also been proposed as sources of ultra high energy cosmic rays~\cite{Mikhailov,Fang}. Although the aforementioned  results don't unequivocally  point to pulsars as sources of TeV energy neutrinos,   there has not been any dedicated searches  for  neutrinos from  pulsars with IceCube.  Therefore, we use the public IceCube neutrino point source catalog to search for a spatial coincidence with individual pulsars.

The outline of this manuscript is as follows. The neutrino and pulsar dataset used in this  work are described in Section~\ref{sec:dataset}. The analysis and results are discussed in Section~\ref{sec:analysis} and we conclude in Section~\ref{sec:conclusions}.

\section{Dataset}
\label{sec:dataset}
The IceCube detector is a neutrino detector located in the South Pole, which detects neutrinos through the Cherenkov light emitted by the  charged leptons travelling through the ice. Its energy threshold is about  100 GeV. 
The IceCube public data release~\cite{IceCubedata} contains both through-going and starting track like events detected between April 2008 and July 2018. This dataset was also used for time-integrated searches for point sources by the IceCube Collaboration~\cite{Icecube}. These track events  primarily consist of charged current interactions of muon or tau neutrinos. The median angular resolution is less than $1^{\circ}$. This  catalog contains 1,134,450 neutrinos and for each neutrino,  its  right ascension (RA), declination ($\delta$),  reconstructed muon energy and   error  in position,   detector zenith and azimuth angle has been made available. 

The pulsar dataset which we use is from v1.68 of  the ATNF catalog and currently consists of 3341 pulsars~\cite{ATNF}\footnote{https://www.atnf.csiro.au/research/pulsar/psrcat/}. For our analysis we only need the right ascension and declination of each pulsar. We do not consider the uncertainty in the pulsar position, since it is much smaller  than the angular error in position of each neutrino. Since our analysis is restricted to searching for neutrino emission from individual pulsars, we do not incorporate the pulsar distance in our analysis. This however would be needed in case of a stacked search.
\section{Analysis}
\label{sec:analysis}
We use the same unbinned maximum likelihood ratio  method as in ~\cite{Hooper,Kamionkowski,LuoZhang,Li22}, which was first proposed in ~\cite{Braun08} and is also used in time-integrated searches done by the IceCube Collaboration~\cite{Icecube}.  For our analysis we  select those neutrino events, with declination within $\pm 5^{\circ}$ of the  pulsar. Most of the pulsars are located close to the galactic plane. Since there are only two pulsars with $\delta> 85^{\circ}$ (85.5$^{\circ}$ and 86.7$^{\circ}$), we have also included the neutrinos within $5^{\circ}$ of  the poles for completeness.
For a dataset of $N$ events, if $n_s$ signal events are attributed to a  pulsar, the probability density of an individual event $i$ is given by:
\begin{equation}
P_i = \frac{n_s}{N} S_i + (1-\frac{n_s}{N}) B_i,
\label{eq1}
\end{equation}
where $S_i$ and $B_i$ represent the signal and background pdfs, respectively.
The likelihood function ($\mathcal{L}$) of the entire dataset, obtained from the product of each individual PDF can be written as:
\begin{equation}
\mathcal{L} (n_s) = \prod_{i=1}^N P_i,
\end{equation}
where $P_i$ is the same as in Eq.~\ref{eq1}. The signal PDF is given by:
\begin{equation}
S_i = \frac{1}{2\pi\sigma_i^2}e^{-(|\theta_i-\theta_s|)^2/2\sigma_i^2}
\label{eq:S}
\end{equation}
where $|\theta_i-\theta_s|$ is the angular distance between the  pulsar and the neutrino, whereas $\sigma_i$ is the angular uncertainty in the neutrino position, expressed in radians. 

\begin{figure}
    \centering
    \includegraphics[scale=0.5]{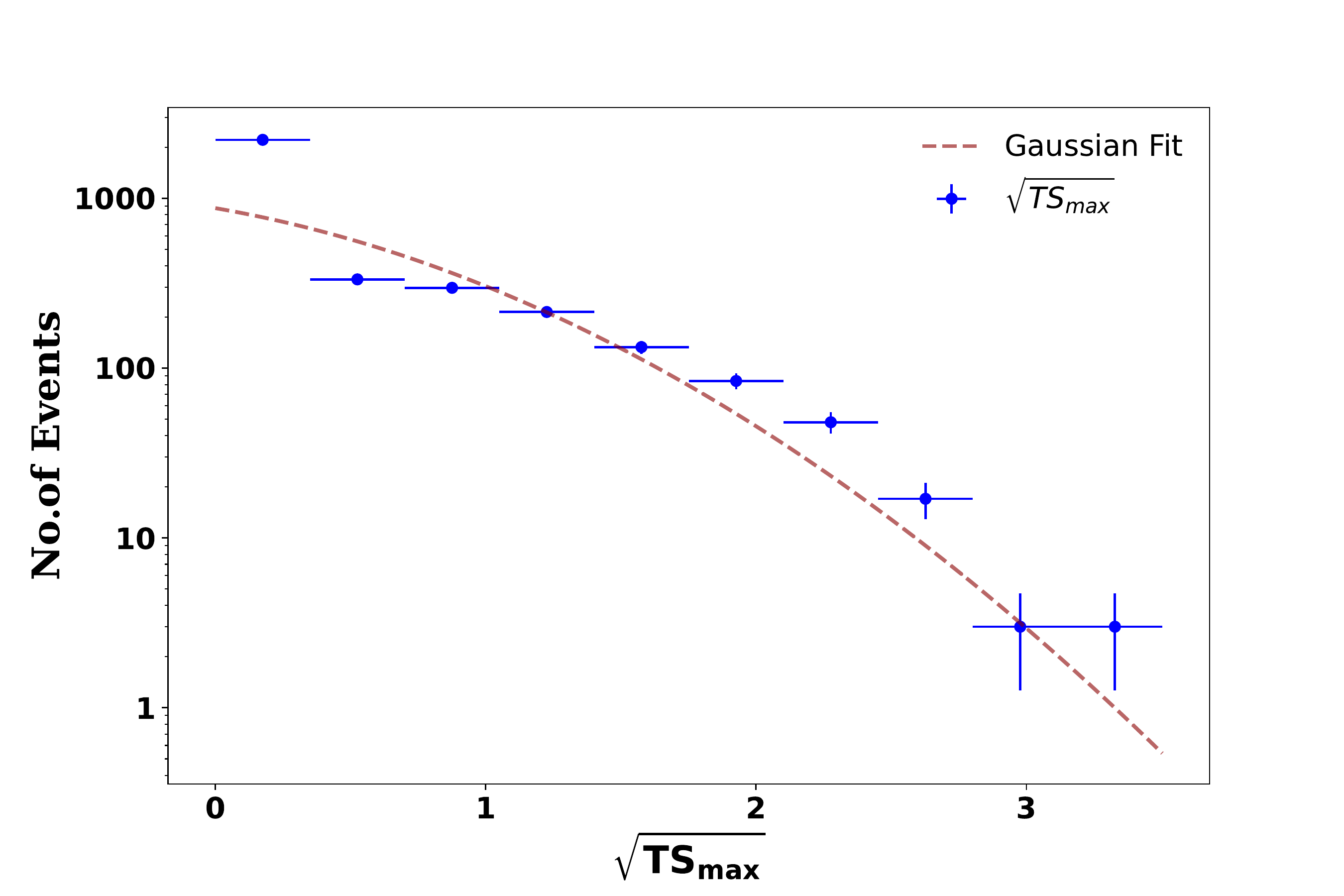}    
    \caption{Histogram of distributions of $\sqrt{TS_{max}}$ along with error bars given by square root of the number of events. The dashed red line shows the best Gaussian fit to this data.}
    \label{fig1}
\end{figure}

\begin{figure}
     \includegraphics[scale=0.4]{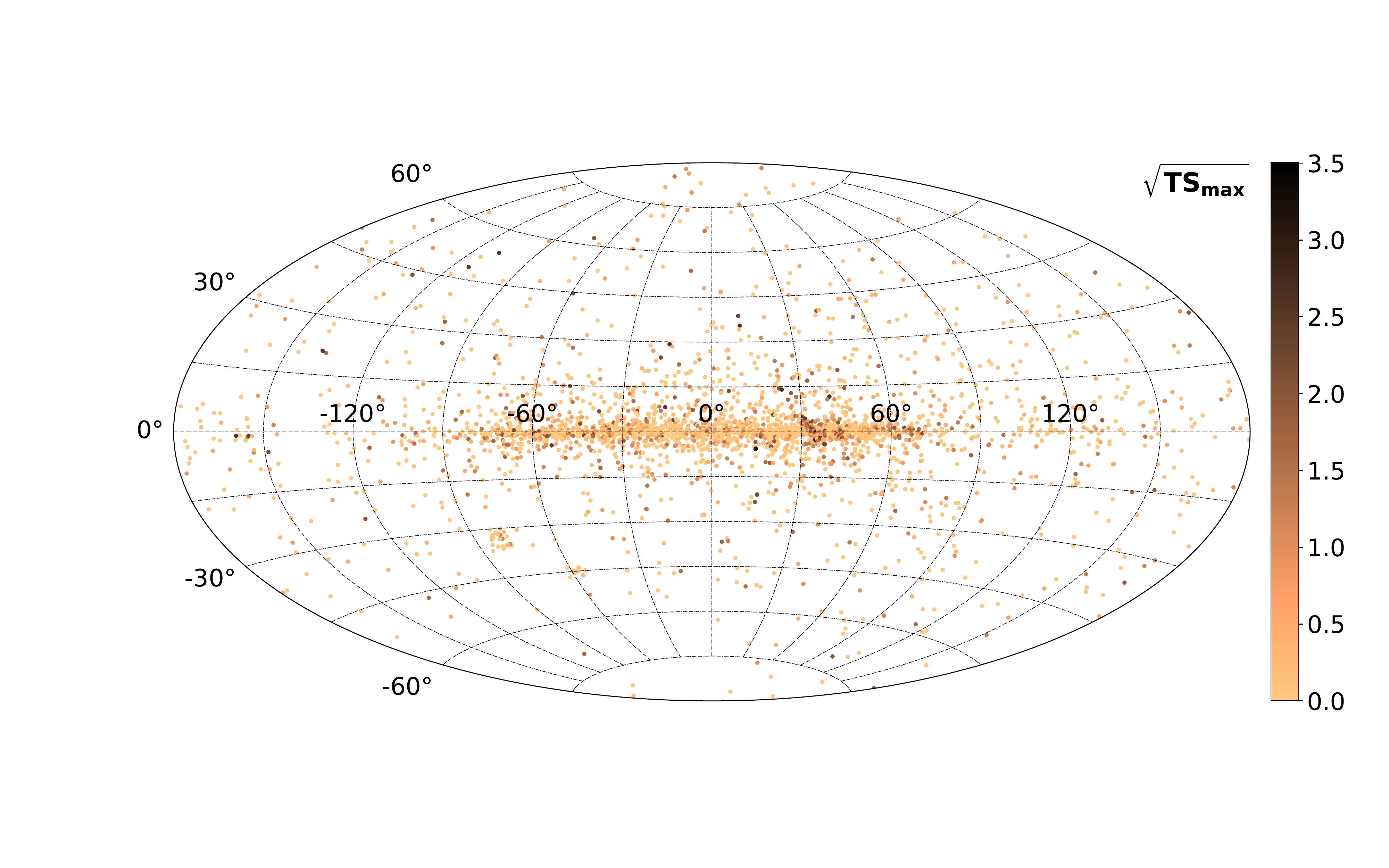}    
    \caption{Skymap distribution of $\sqrt{TS_{max}}$ in Galactic coordinates using Aitoff projection.}
    \label{fig2}
\end{figure}

The background PDF ($B_i$) is determined by the solid angle within  $\delta$ of $\pm 5^{\circ}$ around each pulsar ($\Omega_{\delta \pm 5^{\circ}}$):
\begin{equation}
B_i=\frac{1}{\Omega_{\delta \pm 5^{\circ}}}
\end{equation}
We do not incorporate  the energy information, since the public IceCube catalog only contains the energy of the reconstructed muon and not the energy of the parent neutrino. The detection statistic (or the $Z$-score) used to ascertain the presence of a signal is given by:
\begin{equation}
TS (n_s) = 2 \log \frac{\mathcal{L} (n_s)}{\mathcal{L} (0)}
\end{equation}
 If the null hypothesis is true, $TS (n_s)$ is close to  a  $\chi^2$ distribution for one degree of freedom, according to Wilks' theorem~\cite{Wilks}. The detection significance  is then given by $\sqrt{TS}$.
  We calculated $TS (n_s)$ for all pulsars in the  ATNF catalog. We fit  the histogram of $\sqrt{TS}$ to a Gaussian distribution, where the amplitude is fixed to the ratio of the number of sources to the total number of bins per unit X-axis (similar to ~\cite{Kamionkowski}), while the mean and standard deviation are allowed to vary. The error in each   bin is equal to the square root of the total number of events. This histogram can be found in Fig~\ref{fig1}. Although, a Gaussian distribution is not a good fit to the data because of a slight excess of events with very low value of $TS$ in the first bin, we do not see a statistically significant excess compared to the background.  This excess could also be due to the fact that the PDF of 
  $TS (n_s)$ is a superposition of $\chi^2$ and $\delta$ function, since $n_s$ is close to the physical boundary~\cite{Wolf}. 
  The  excess seen between  $\sim 1.5-2.5$ can be explained by inaccuracies in the PDF, as argued in~\cite{Hooper}, and also by  incorporating the trial factors. The skymap distribution of this statistic in galactic coordinates can be found in Fig.~\ref{fig2}.  Therefore, we conclude that none of the current pulsars in the ATNF catalog (on their own) contribute to the diffuse neutrino flux seen in IceCube. We show the pulsars with the  three highest $TS_{max}$ values in Table~\ref{table1}.

\begin{center}
\begin{table}[ht!]
\setlength{\arrayrulewidth}{0.3mm}
\setlength{\tabcolsep}{18pt}
\begin{tabular}{ P{1.75cm} P{1.5cm} P{1.5cm}}
	\hline\hline
	\centering
	\large Pulsar & \large $\tilde{n}_s$ & \large $TS_{max}(\tilde{n}_s)$\\
	\hline
	J1849+0037g & 43.5 & 12.3\\
	
	J1707-4341 & 17.4 & 11.8\\

	J1838-1849 & 16.5 & 10.4\\
	\hline\hline
\end{tabular}
\caption{Pulsars with highest $TS_{max}(\tilde{n}_s)$ values}
\label{table1}
\end{table}
\end{center}

\section{Conclusions}
\label{sec:conclusions}
In this work we look for a spatially significant association between  individual radio pulsars  and TeV energy neutrinos detected by IceCube using the publicly available point source neutrino catalog, which consists of  11,34,450 neutrinos obtained from the  muon track data using 10 years of observations from 2008-2018. For our analysis, we used all the 3341 radio pulsars compiled in the  latest version of the  ATNF catalog. The analysis was done using the unbinned maximum likelihood ratio method~\cite{Braun08}, which has also been used in astrophysical point-source searches within and outside the  IceCube collaboration. 

The distribution of the detection significance for each of the pulsars is shown in Fig.~\ref{fig1} and the skymap distribution in galactic coordinates in Fig.~\ref{fig2}. When we fit the histogram of the detection significance to a Gaussian distribution, we find a slight excess of events in the first bin. However, the distribution of events at large significance is consistent with background, indicating that that there is no statistically significant excess.
We should point that PSR B1509-58 which showed a $2.6\sigma$ excess, when one combines the data from Super-K and MACRO~\cite{Desai22}, has a $TS_{max}$
value equal to 0, implying that no excess is seen from this pulsar in IceCube data.
Therefore, we can rule out any of the known pulsars as sources of  TeV energy high energy neutrinos detected in IceCube.

We should point out that our current analysis  does not include a stacked contribution from all the pulsars, which we shall defer to a future work.

\section*{Acknowledgments}
We are grateful to Bei Zhou, Jiawei Luo, Rong-Lan Li, and Dan Hooper for patiently explaining the technical details of their works. and to Ranjan Laha for comments and useful feedback  on our manuscript. We are also grateful to the IceCube collaboration for making their point source neutrino catalog publicly available.


\bibliography{main}
\end{document}